# Investigation of Network Architecture for Multimodal Head-and-Neck Tumor Segmentation

Ye Li
CAMCA
MGH/HMS
Boston, USA
gary.li@mgh.harvard.edu

Junyu Chen
Department of Electrical and Computer Engineering
Johns Hopkin University
Baltimore, USA
jchen245@jhmi.edu

Se-in Jang
CAMCA
MGH/HMS
Boston, USA
sjang7@mgh.harvard.edu

Kuang Gong
CAMCA
MGH/HMS
Boston, USA
kgong@mgh.harvard.edu

Quanzheng Li
CAMCA
MGH/HMS
Boston, USA
li.quanzheng@mgh.harvard.edu

*Abstract*— Inspired by the recent success of Transformers for Natural Language Processing and vision Transformer[1] for Computer Vision, many researchers in the medical imaging community have flocked to Transformer-based networks for various main stream medical tasks such as classification, segmentation, and estimation. In this study, we analyze, two recently published Transformer-based network architectures for the task of multimodal head-and-tumor segmentation and compare their performance to the de facto standard 3D segmentation network – the nnU-Net. Our results showed that modelling long-range dependencies may be helpful in cases where large structures are present and/or large field of view is needed. However, for small structures such as head-and-neck tumor, the convolution-based U-Net architecture seemed to perform well, especially when training dataset is small and computational resource is limited.

## I. Introduction

This paper intends to analyze, in the context of multimodal tumor segmentation, the capabilities of the two recent and major segmentation network architectures, namely Transformer-based U-Net [2, 3] and nnU-Net[4]. At first glance, it may be thought that the capability of capturing long-range dependencies brought by the Transformer-based models is a pure gain to the U-Net models. Certainly, the self-attention modules are effective at modeling the long-range dependencies but they come at the cost of a sharp increase in the number of parameters in the network. In theory, more data and computational power is needed to train such a network. Besides, it has been shown in [5, 6] that Transformers require a considerable amount of data to achieve superior performance due to their weak inductive bias. Thus, it may be more appropriate to use a smaller network when data is limited. Aside from the network architecture, it is also worth noting the importance of the specific goal of the segmentation task, i.e., whether the segmentation target is a large or small structure and long-range dependency modeling is or not really needed. For small targets, the information needed to segment the target is more available in the local/neighboring areas around the target than in the areas that are far away from it. In such cases, modeling of long-range dependencies may not be necessary as information that is far away from the target would never be used to segment the target. Choosing a large Transformer model for segmenting a small target would be not shrewd as it comes with higher data demand and computational cost.

In this paper, we aim to provide a side-by-side comparison for the performance as well as computational demand for two recent Transformer-based U-Net models (Swin UNETR[2] and UNETR [3]) and the nnU-Net segmentation model, in the context of tumor segmentation. Specifically, we applied Swin UNETR, UNETR and nnU-Net on two public datasets, namely the HECKTOR 2021 dataset and a TCIA dataset. We compared their performance in the segmentation of head-and-neck primary tumor (a small structure), as measured by the Dice coefficient, as well as the network parameters, and GPU times. In addition, we analyzed results of these three networks on the BraTS 2021 dataset, which contains brain tumor of three different sizes. Lastly, we analyzed the tradeoff among the dataset size, segmentation target of various sizes, and network parameters for Transformer-based U-Net and the U-Net architecture.

## II. Method

### A. Studied Network Architectures

To study the tradeoffs mentioned above, we investigated three leading network architectures for medical image segmentation on the task of segmenting head-and-neck primary tumors from FDG-PET and CT images. The network architectures are shown in Fig. 1.

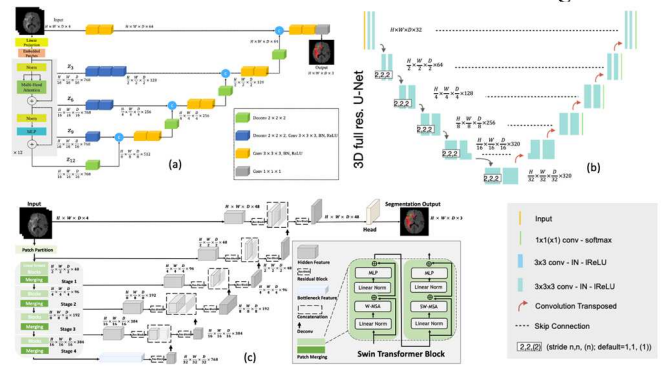

Figure 1. Three network architectures investigated in the study. (a) UNETR (b) nnU-Net 3D full resolution and (c) SWINUNETR

## B. Datasets

Two public datasets for Head and Neck tumor segmentation were used in the study: the HECTOR challenge dataset and the TCIA HNC dataset. For these datasets, the primary gross tumor volume (GTVt) for these patients were annotated by expert radiologists (HECTOR) and histological data (TCIA). The HECTOR challenge training dataset includes 224 cases (of which 44 were used for validation), each with two modalities: a) PET and b) CT where are rigidly aligned, and resampled to a $1 \times 1 \times 1$ mm isotropic resolution. The HECTOR data was preprocessed using the codes provided by the challenge website. The input image size for the HECTOR dataset is $144 \times 144 \times 144$. The TCIA dataset is consisted of 122 cases (of which 24 were used for validation), with the same imaging modalities and images resampled to the same isotropic resolution. The input image size for the TCIA dataset is $128 \times 128 \times 128$.

## C. Implementation Details

nnU-Net and UNETR were implemented using codes from the official GitHub repository. Swin UNETR was implemented using PyTorch and MONAI. All networks were trained on a DGX-1 cluster with 4 NIVIA V100 GPUs. Table 1 details the configurations of UNETR and Swin UNETR architecture and number of parameters.

**Table 1.** Summary of UNETR, Swin UNETR, and nnU-Net hyperparameters

| | Embed Dimension | Feature Size | Number of Blocks | Window Size | Number of Heads | Parameters |
|---|---|---|---|---|---|---|
| SWINU NETR | 768 | 48 | [2,2,2,2] | [7,7,7] | [3,6,12,24] | **138M** |
| | Embed Dimension | Feature Size | Number of Stages | MLP dimension | Number of Heads | Parameters |
| UNETR | 768 | 32 | 5 | 3072 | 12 | **104M** |
| | Mode | Conv Kernel Size | Number of Stages | Pooling Kernel Size | Trainer | Parameters |
| nnU-Net | 3D full resolution | [3,3,3] | 5 | [2,2,2] | nnUNetTrainerV2 | **16.2M** |

The learning rate was set to 0.0001. All input images (both PET and CT) were scaled to the intensity range of [0,1]. Random patches of $96 \times 96 \times 96$ were cropped from 3D image volumes during training. In addition, random axis mirror flips with a probability of 0.2 was applied for all 3 axes. We also applied data augmentation transforms of random per channel intensity shift in the range of [-0.1, 0.1], and random scale of intensity in the range of [0.9, 1.1] to input image channels. The batch size per GPU was set to 10 and 1 for UNETR and Swin UNTER, respectively. All models were trained for a total of 2000 epochs with a linear warmup and using a cosine annealing learning rate scheduler. A sliding window approach with an overlapping of 0.7 for neighboring voxels was adopted for inference.

## III. RESULTS AND DISCUSSION

The quantitative model performance, as measured by the mean Dice score, showed that the nnU-Net was the best among the three models tested in this study, for the task of head-and-neck primary tumor segmentation in PET/CT images. In [2], the author also reported very minor (0.004) dice improvement compared to nnU-Net for the smallest tumor using the BraTS chanllenge dataset. Together, these results indicated that, for small structure segmentation, it might be beneficial to stick with the smaller sized convolution-based U-Net architecture, especially when training dataset is small and computational resource is limited.

**Table 2.** Summary of mean Dice score values on the validation dataset.

| Dice Score | UNETR | Swin UNETR | nnU-Net |
|---|---|---|---|
| TCIA | 0.708±0.088 | 0.741±0.083 | **0.765±0.082** |
| HECTOR | 0.693±0.250 | 0.733±0.191 | **0.762±0.154** |

In Fig. 2, we showed two representative cases to demonstrate the advantage of U-Net as well as Transformer on a case-by-case scenario. The images in the first row showed that the U-Net outperformed the Transformer-based models in segmenting single, isolated tumor. In the second row, the Transformer-based models demonstrated superior performance to nnU-Net in segmenting tumor which has more connected area, potentially due to modeling of long-range dependencies brought by the Transformer encoder.

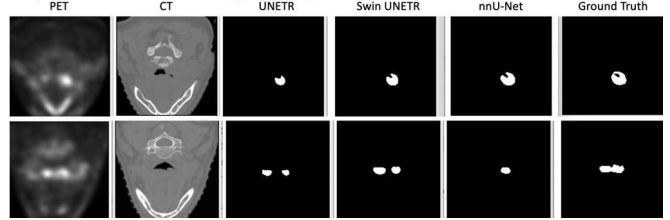

Figure 2. From left to right are input PET image, CT image, inferenced mask from UNETR, SWINUNETR, nnU-NET, and ground truth.

## IV. CONCLUSION

This study investigated the tradeoff between network size and performance gain for three leading network architectures for multi-modal semantic segmentation. Our results showed that modelling long-range dependencies may be helpful in cases where large structures are present and/or large field of view is needed. However, for small structures, the convolution-based U-Net architecture seemed to perform well, especially when training dataset is small and computational resource is limited.